\documentstyle[12pt]{ioplppt} 

\eqnobysec

\ifx\fiverm\undefined
	\newfont\fiverm{cmr5}
\fi
\input prepictex
\input pictex
\input postpictex

\begin{document} 
\title{Ground state and low excitations of an
integrable chain with alternating spins}[XXZ($\frac{1}{2},1$)
Heisenberg spin chain] \author{St Mei\ss ner\footnote{email:
meissner@qft2.physik.hu-berlin.de} and B - D D\"orfel\footnote{email:
doerfel@ifh.de}} \address{Institut f\"ur Physik,
Humboldt-Universit\"at , Theorie der Elementarteilchen\\
Invalidenstra\ss e 110, 10115 Berlin, Germany}

\begin{abstract} 
An anisotropic integrable spin chain, consisting of
spins $s=1$ and $s=\frac{1}{2}$, is investigated \cite{devega}. It is
characterized by two real parameters $\bar{c}$ and $\tilde{c}$, the
coupling constants of the spin interactions. For the case $\bar{c}<0$
and $\tilde{c}<0$ the ground state configuration is obtained by means
of thermodynamic Bethe ansatz. Furthermore the low excitations are
calculated. It turns out, that apart from free magnon states being
the holes in the ground state rapidity distribution, there exist
bound states given by special string solutions of Bethe ansatz
equations (BAE) in analogy to \cite{babelon}. The dispersion law of
these excitations is calculated numerically.  
\end{abstract} 


\maketitle 
\section{Introduction} Since the development of the quantum
inverse scattering method (QISM) \cite{fad,kul} many generalisations
of the well known XXZ($\frac{1}{2}$) model have been investigated
\cite{takh,babu,tsve}. The appearance of interesting new features of
these models made those analysises very fruitful.

In 1992 de Vega and Woynarovich suggested a new kind of
generalisation by constructing a model containing
spin-$\frac{1}{2}$- and spin-$1$-particles \cite{devega}. In comparison with
models containing only one kind of spin it shows a richer structure due to 
additional parameters, the spin interaction coupling constants, making an
investigation worthwhile.

Its isotropic limit XXX($\frac{1}{2},1$) has been studied in papers 
\cite{devega1,devega2}. The generalisation procedure was extended to pairs of
higher spins ($\frac{1}{2},S$) and ($S',S$) in \cite{aladim1,aladim2,martins}.
Except \cite{devega1} only the conformally invariant case with positive 
coupling was considered. Our aim therefore is to go beyond that point. We 
intend to continue along this line in a subsequent publication.

In this paper we want to study the XXZ($\frac{1}{2},1$) model with strictly 
alternating spins. Definitions are reviewed in section 2. In section 3 we 
carry out the thermodynamic Bethe ansatz (TBA) and obtain the ground state for
negative coupling of the spin interactions. Section 4 deals with the higher 
level Bethe ansatz for low excitations above this ground state and section 5 
contains our conclusions.
\section{Description of the model}
We consider the Hamiltonian of a spin chain of length $2N$ \cite{devega}
\begin{equation}\label{ham}
{\cal H}(\gamma) = \bar{c} \bar{\cal H}(\gamma) + \tilde{c} \tilde{\cal H}
(\gamma) - H S^z,
\end{equation}
where $\bar{\cal H}$ couples two spins $s=\frac{1}{2}$ and one spin $s=1$, 
while the converse is true for $\tilde{\cal H}$ (for explicit construction see
again \cite{devega}, bars and tildes are interchanged with respect to this 
reference as in subsequent publications of de Vega {\it et al.}). 
$H$ is an external magnetic field. We impose periodic boundary conditions.

The Hamiltonian contains an XXZ type anisotropy, which is parametrized by 
$e^{i\gamma}$ or $e^{-\gamma}$ respectively. We investigate the weak 
antiferromagnetic case, i.e. the parametrization is $e^{i\gamma}$ and we 
restrict ourselves to $0<\gamma<\pi/2$. Moreover we have two additional real 
parameters $\tilde{c}$ and $\bar{c}$ being the coupling constants 
of the spin interactions and dominating the qualitative behaviour of the 
model.

The Bethe ansatz equations (BAE) determining the solution of the model are
\begin{equation}\label{bae}
\fl \left( \frac{\sinh(\lambda_j+i\frac{\gamma}{2})}{\sinh(\lambda_j-i
\frac{\gamma}{2})}
\frac{\sinh(\lambda_j+i\gamma)}{\sinh(\lambda_j-i\gamma)} \right)^N =
-\prod_{k=1}^{M}\frac{\sinh(\lambda_j-\lambda_k+i\gamma)}{\sinh(\lambda_j-
\lambda_k-i\gamma)},\qquad j=1\dots M.
\end{equation}
One can express energy, momentum and spin in terms of BAE roots $\lambda_j$:
\begin{eqnarray}\label{en}
E = \bar{c} \bar{E} + \tilde{c} \tilde{E} - \left( \frac{3N}{2} - M \right) H,
\nonumber\\
\bar{E} = - \sum_{j=1}^{M} \frac{2\sin\gamma}
{\cosh2\lambda_j - \cos\gamma},
\nonumber\\
\tilde{E} = - \sum_{j=1}^{M} 
\frac{2\sin2\gamma}{\cosh2\lambda_j - \cos2\gamma},
\end{eqnarray}
\begin{equation}\label{mom}
P =\frac{i}{2}\sum_{j=1}^{M} \left\{ \log \left(\frac{\sinh(\lambda_j+i\frac{
\gamma}{2})}{\sinh(\lambda_j-i\frac{\gamma}{2})} \right) + 
\log \left( \frac{\sinh(\lambda_j+i\gamma)}{\sinh(\lambda_j-i\gamma)} \right) 
\right\},
\end{equation}
\begin{equation}\label{spin}
S_z = \frac{3N}{2} - M.
\end{equation}
Here we have substracted a constant in the momentum in order to make 
this magnitude vanishing for the ferromagnetic state. Furthermore additive
constants in $\bar{\cal H}$ and $\tilde{\cal H}$ are dropped due to different
normalisation compared with \cite{devega}.

\section{Thermodynamic Bethe ansatz (TBA) and ground state for negative 
coupling constants}
We want to study the model in the thermodynamic limit $N\to\infty$. Then only 
solutions of the string type occur (for $M$ fix):
\begin{equation}\label{string}
\lambda_{\alpha}^{n,j,\nu} = \lambda_{\alpha}^{n,\nu} + i(n+1-2j)\frac{\gamma}
{2} + \frac{1}{4} i \pi (1-\nu) + \delta_{\alpha}^{n,j,\nu},\qquad j=1\dots n.
\end{equation}
Here $\lambda_{\alpha}^{n,\nu}$ is the real center of the string, $n$ is the
string length and $\nu$ the parity of the string with values $\pm1$. The last
term is a correction due to finite size effects.

Now the permitted values of the string length $n$ and related parities $\nu$ 
depending on the anisotropy $\gamma$ are to determine. Several 
approaches have been used to obtain them for the massive Thirring model 
\cite{kor} and for the XXZ($\frac{1}{2}$) \cite{taka} and XXZ(S) models 
\cite{kir1,kir2,frahm} all leading to the Takahashi conditions, a 
system of inequalities describing admissible string length and parities: 
\begin{equation}\label{cond}
\nu_n \sin \gamma j \sin \gamma (n-j) > 0, \qquad j=1 \dots n-1.
\end{equation}
In the XXZ(S) case additional restrictions on the spin arise.

By applying similar methods to our model one can show that fulfilling the 
Takahashi conditions is necessary and sufficient for an admissible pair
($n,\nu_n$) for the number of magnons $M$ fix. On the other hand we did not 
succeed in deriving such conditions in the case $N,M\to\infty,M/N$ fix from the
BAE directly. Nevertheless there is a series of arguments favouring the
conjecture that in this case the same strings as in the previous one can exist.
In the following we want to state a few of them briefly. 

Generally the string hypothesis is believed to be valid in the case 
$N,M\to\infty,M/N$ fix (i.e. string corrections are exponentially small) for 
non-vanishing external magnetic field (see e.g. \cite{wieg} and 
\cite{devega2}). For zero magnetic field the numerical results of Frahm 
{\it et al.} show that string corrections are of orders $O(1/N)$ to $O(1)$, 
what does not affect the thermodynamics of the model.

For further discussion it is important, that, as already proven in 
\cite{devega}, the 'sea-strings' are exponentially exact.

Furthermore Takahashi and Suzuki  obtained a phenomenological rule determining 
admissible strings for the XXZ($\frac{1}{2}$) model proven to be 
equivalent to the Takahashi conditions. Fulfilling this criterion 
requires that a singlet state can be constructed by introducing a maximum 
number of strings of one type. Kirillov and Reshetikhin generalized this 
criterion naturally to higher spins. In the XXZ(S) it is as well a 
selection rule for permitted spin values as a criterion for an admissible pair 
($n,\nu_n$). 

In the following we therefore shall assume the existence of strings fulfilling 
the Takahashi conditions.

Substituting (\ref{string}) into (\ref{bae}) and taking the logarithm yields
\begin{eqnarray}\label{logbae}
N t_{j,1}(\lambda_{\alpha}^{n_j}) + N t_{j,2}(\lambda_{\alpha}^{n_j}) 
= 2\pi I_{\alpha}^{n_j} + \sum_{k} \sum_{\beta} \Theta_{jk}(\lambda_{\alpha}
^{n_j}-\lambda_{\beta}^{n_k},\nu_j\nu_k)
\end{eqnarray}
with the known notations
\begin{equation}
t_{j,2S}(\lambda) = \sum_{k=1}^{\mbox{min}(n_j,2S)} 
f(\lambda,|n_j-2S|+2k-1,\nu_j),
\end{equation}
\begin{eqnarray}
\fl
\Theta_{jk}(\lambda) = f(\lambda,|n_j-n_k|,\nu_j\nu_k) 
\nonumber\\
+ f(\lambda,(n_j+n_k),\nu_j\nu_k) 
+ 2 \!\!\!\!\!\!\!\sum_{k=1}^{min(n_j,n_k)-1} 
\!\!\!\!f(\lambda,|n_j-n_k|+2k,\nu_j\nu_k),
\end{eqnarray}
and
\begin{eqnarray}
f(\lambda,n,\nu) = \left\{ \begin{array}{r@{\quad\quad}l}
					0 & n\gamma/\pi \in \bold Z\\
					2\nu \arctan ((\cot (n\gamma/2))^{\nu}
				\tanh\lambda) & n\gamma/\pi \notin \bold Z
			  \end{array} \right. .
\end{eqnarray}
Here we have used that a given string length $n>1$ corresponds to a unique 
parity, what is a consequence of (\ref{cond}). The numbers $I_{\alpha}^{n_j}$ 
are half-odd-integers counting the strings of length $n_j$.

Introducing particle and hole densities in the usual way (see e.g. \cite{babu} 
eqs. 60-63) we perform the limiting process $N\to\infty$
\begin{eqnarray}\label{intbae}
a_{j,1}(\lambda) + a_{j,2}(\lambda)
= (\rho_j(\lambda)+\tilde{\rho}_j(\lambda))(-1)^{r(j)}+
\sum_{k} T_{jk} * \rho_k(\lambda),
\end{eqnarray}
where $a*b(\lambda)$ denotes the convolution
\begin{equation}
a*b(\lambda) = \int_{-\infty}^{\infty}d\mu a(\lambda-\mu)b(\mu)
\end{equation}
and
\begin{equation}
a_{j,2S}(\lambda) = \frac{1}{2\pi}\frac{d}{d\lambda}t_{j,2S}(\lambda),
\qquad
T_{j,k}(\lambda) =  \frac{1}{2\pi}\frac{d}{d\lambda}\Theta_{j,k}(\lambda).
\end{equation}

The sign $(-1)^{r(j)}$ results from the requirement of positive densities in 
the 'non-interacting' limit (i.e. only one string type is present)\cite{frahm}.

The analysis of (\ref{cond}) remains complicated for arbitrary $\gamma$ 
\cite{taka}, but at special values the picture becomes easy, namely
$\gamma=\pi/\mu,\quad\mu\dots$ integer, $\mu\geq 3$. Following Babujian and 
Tsvelick \cite{tsve} we want to consider that case only.

Due to periodicity we are left with strings
\begin{enumerate}
\item
$n_j=j,\qquad j=1\dots\mu-1,\qquad\nu_j=1$,
\item
$n_{\mu}=1,\qquad\nu_{\mu}=-1$.
\end{enumerate}
Equation (\ref{intbae}) then reduces to
\begin{eqnarray}
a_{j,1}(\lambda) + a_{j,2}(\lambda)
=
\rho_j(\lambda) + \tilde{\rho}_j(\lambda) + \sum_{k=1}^{\mu} T_{jk}*\rho_k
(\lambda), \qquad j=1 \dots \mu-1,
\nonumber\\
a_{\mu,1}(\lambda) + a_{\mu,2}(\lambda)
=
-\rho_{\mu}(\lambda) -\tilde{\rho}_{\mu}(\lambda) + \sum_{k=1}^{\mu} T_{\mu k}*
\rho_k(\lambda).
\end{eqnarray}

We are now able to express energy, momentum and spin in terms of the densities 
via (\ref{en}), (\ref{mom}) and (\ref{spin}). The standard procedure leads to 
equations determining the equilibrium state at temperature $T$:
\begin{eqnarray}\label{tbae}
\fl
T\ln\left(1+\exp\left(\frac{\epsilon_j}{T}\right)\right) = 
\nonumber\\
H n_j
-2\pi\bar{c}a_{j,1}(\lambda) - 2\pi\tilde{c} a_{j,2}(\lambda)
+ \sum_{k=1}^{\mu} T\ln\left(1+\exp\left(\frac{-\epsilon_k}{T}\right)\right) 
* A_{jk}(\lambda)
\end{eqnarray}
with
\begin{equation}
A_{jk}(\lambda)=(-1)^{r(k)}T_{jk}(\lambda,\nu_j\nu_k) + \delta(\lambda)
\delta_{jk}
\end{equation}
and
\begin{equation}
\frac{\tilde{\rho}_j}{\rho_j}=\exp\left(\frac{\epsilon_j}{T}\right).
\end{equation}
Again the free energy can be expressed in terms of our new variables 
$\epsilon_j(\lambda)$:
\begin{equation}\label{free}
\fl
2 {\cal F} = \frac{F}{N} = -\int_{-\infty}^{\infty} d\lambda \sum_{j=1}^{\mu}
(-1)^{r(j)}(a_{j,1}(\lambda) + a_{j,2}(\lambda))T\ln\left(1+\exp\left(\frac
{-\epsilon_j}{T}\right)\right)  - \frac{3H}{2}.
\end{equation}
Reversing the operator $A_{jk}$ in (\ref{tbae}) by applying 
\begin{equation}
C_{jk} = \delta(\lambda)\delta_{jk} - s(\lambda)(\delta_{j+1k}+\delta_{j-1k}),
\qquad j,k=1\dots\mu
\end{equation}
with
\begin{equation}
s(\lambda) = \frac{1}{2\gamma\cosh(\pi\lambda/\gamma)}
\end{equation}
yields
\begin{eqnarray}\label{invtbae}
\epsilon_1(\lambda) &=& Ts*\ln(f(\epsilon_2)) - 2\pi\bar{c} s(\lambda),
\nonumber\\
\epsilon_2(\lambda) &=& Ts*\ln(f(\epsilon_3)f(\epsilon_1)) 
- 2\pi\tilde{c} s(\lambda),
\nonumber\\
\epsilon_j(\lambda) &=& Ts*\ln(f(\epsilon_{j+1})f(\epsilon_{j-1})) 
+ \delta_{j\mu-2}
Ts*\ln(f(-\epsilon_{\mu})),j=3\dots\mu-2,
\nonumber\\
\epsilon_{\mu-1}(\lambda) &=&\frac{H\mu}{2} + Ts*\ln(f(\epsilon_{\mu-2})),
\nonumber\\
\epsilon_{\mu}(\lambda) &=& \frac{H\mu}{2} - Ts*\ln(f(\epsilon_{\mu-2})),
\end{eqnarray}
with
\begin{eqnarray}
f(x)=1+e^{x/T}.
\end{eqnarray}
From these equations it can be easily established that $\epsilon_j\geq0$ for 
$j=3 \dots\mu-1$.

Since we are interested in the ground state we take the limit $T\to0$ in 
equations (\ref{tbae}) and (\ref{free}). Taking into account that 
$\epsilon_j\geq0$ for $j=3\dots\mu-1$ we have
\begin{equation}\label{zerotbae}
\fl
\epsilon_j^+(\lambda) = H n_j -2\pi\bar{c}a_{j,1}(\lambda) - 2\pi\tilde{c} 
a_{j,2}(\lambda)
- \epsilon_1^- * A_{j1}(\lambda) - \epsilon_2^- * A_{j2}(\lambda)
- \epsilon_{\mu}^- * A_{j\mu}(\lambda),
\end{equation}
\begin{eqnarray}\label{zerotfree}
\fl
2 {\cal F} = \frac{F}{N} = \int_{-\infty}^{\infty} d\lambda \left[ (a_{1,1}
(\lambda) + a_{1,2}(\lambda))\epsilon_1^-(\lambda)  
\right.
\nonumber\\
\left.
+ (a_{2,1}(\lambda) + a_{2,2}(\lambda))\epsilon_2^-(\lambda)
- (a_{\mu,1}(\lambda) + a_{\mu,2}(\lambda))\epsilon_{\mu}^-(\lambda) \right] 
- \frac{3H}{2},
\end{eqnarray}
where $\epsilon_j^+$ and $\epsilon_j^-$ denote positive and negative parts of 
the function $\epsilon_j$ respectively.

Now we want to discuss equation (\ref{zerotbae}) for the case $H=0^+$ in 
dependence of the signs of $\bar{c}$ and $\tilde{c}$.
\begin{enumerate}
\item
$\bar{c}>0,\tilde{c}>0$

This sector has been investigated completely by de Vega and Woynarovich 
\cite{devega}. The solution is
\begin{equation}
\epsilon_j(\lambda) = -2\pi\bar{c}s(\lambda)\delta_{j1} 
-2\pi\tilde{c}s(\lambda)\delta_{j2}.
\end{equation}
\item
$\bar{c}>0,\tilde{c}<0$

The picture in this sector is difficult, since the quadrant is divided by a 
phase line starting in the origin and going to infinity, which we obtained from
the TBA equations. In the lower area the model behaves as in case (iv), while
in the upper one we expect for the ground state a mixture of strings of length 
$1$ with different parities and finite Fermi zones. We hope to return to this 
case in greater detail later.
\item
$\bar{c}<0,\tilde{c}>0$

The situation in this sector is similar to the case above. Again we have a 
phase line producing one area of case (iv) type and another one where the 
ground state is expected to be formed by the $(2,+)$- and $(1,-)$-strings 
having again finite Fermi zones.
\item
$\bar{c}\leq0,\tilde{c}\leq0$

In this case $\epsilon_1(\lambda)$ and $\epsilon_2(\lambda)$ are positive. 
The only string present in the ground state is $(1,-)$. 
Equations (\ref{zerotbae}) and (\ref{zerotfree}) reduce to 
\begin{equation}\label{zerotbae1}
\epsilon_j^+(\lambda) = -2\pi\bar{c}a_{j,1}(\lambda) 
- 2\pi\tilde{c} a_{j,2}(\lambda) - \epsilon_{\mu}^- * A_{j\mu}(\lambda),
\end{equation}
\begin{equation}\label{zerotfree1}
2 {\cal F} = \frac{F}{N} = - \int_{-\infty}^{\infty} d\lambda (a_{\mu,1}
(\lambda) + a_{\mu,2}(\lambda))\epsilon_{\mu}^-(\lambda).
\end{equation}
The solutions of (\ref{zerotbae1}) can be given via their Fourier transforms:
\begin{eqnarray}
\fl\hat{\epsilon}_{1}(p) =-2\pi\bar{c}\frac{\cosh[p\gamma(\mu-2)/2]}
{\cosh[p\gamma(\mu-1)/2]}
-2\pi\tilde{c}\frac{\cosh[p\gamma(\mu-3)/2]}{\cosh[p\gamma(\mu-1)/2]},
\nonumber\\
\fl\hat{\epsilon}_{j}(p) =-2\pi\bar{c}\frac{\cosh[p\gamma(\mu-j-1)/2]}
{\cosh[p\gamma(\mu-1)/2]}
-2\pi\tilde{c}\frac{2\cosh(p\gamma/2)\cosh[p\gamma(\mu-j-1)/2]}{\cosh[p\gamma
(\mu-1)/2]},
\nonumber\\
\fl\hat{\epsilon}_{\mu-1}(p) = -2\pi\bar{c}\frac{1}{2\cosh[p\gamma(\mu-1)/2]} 
-2\pi\tilde{c}\frac{\cosh(p\gamma/2)}{\cosh[p\gamma(\mu-1)/2]},
\nonumber\\
\fl\hat{\epsilon}_{\mu}(p) = 2\pi\bar{c}\frac{1}{2\cosh[p\gamma(\mu-1)/2]} 
+2\pi\tilde{c}\frac{\cosh(p\gamma/2)}{\cosh[p\gamma(\mu-1)/2]}.
\end{eqnarray}
The ground state energy is
\begin{eqnarray}
\fl 2 {\cal F} = \frac{F}{N} = \bar{c} \int_{-\infty}^{\infty} dp 
\frac{\sinh(p\gamma/2)+\sinh(p\gamma)}{2\cosh[p\gamma(\mu-1)/2]\sinh(p\gamma
\mu/2)} 
\nonumber\\
+ \tilde{c} \int_{-\infty}^{\infty} dp 
\frac{[\sinh(p\gamma/2)+\sinh(p\gamma)]\cosh(p\gamma/2)}{\cosh[p\gamma(\mu-1)
/2]\sinh(p\gamma\mu/2)}.
\end{eqnarray}
For $\bar{c}=\tilde{c}$ the continuum limit provides a conformal invariant 
theory. Due to the existence of {\it one} type of elementary excitations the
central charge of the Virasoro algebra is $c=1$. That is remarkable, because 
in sector (i) $c=2$ \cite{devega}. The point $\bar{c}=\tilde{c}=0$ must be 
singular then, when passed on the conformal line.
\item
$\bar{c}=0,\tilde{c}>0$

\item[(vi)]
$\bar{c}>0,\tilde{c}=0$

Both cases have been briefly considered in [14]. The two lines are infinitely
high singular, because of the infinite degeneration of the ground state.
\end{enumerate}
In the following we shall consider the case (iv) only.

\section{Higher level Bethe ansatz for low excitations}
In this section we want to derive equations for excitations above the ground 
state in the case $\bar{c}<0,\tilde{c}<0$. Starting point of our analysis is
the result of section 3. Though we have derived the ground state configuration
of BAE roots only for the special case $\gamma=\pi/\mu,\mu$ integer,
we extend this result to the whole range of $\gamma,0<\gamma<\pi/2$, 
motivated by the results of Frahm {\it et al.} \cite{frahm}, who
found for the XXZ(S) Hamiltonians, where only special intervals of the
anisotropy $\gamma$ are permitted, that the ground state configuration does not
change within any of these intervals. Since in our case no restrictions on
$\gamma$ arise, i.e. we have only one permitted interval, we expect the ground 
state configuration to be the one obtained for the above mentioned special 
values in the whole $\gamma$-range.

We write down the BAE for this ground state in the limit $N\to\infty$ with a 
finite number of excitations (holes in the ground state rapidity distribution
and additional complex roots):
\begin{eqnarray}\label{logbae1}
\fl \bar{\phi}'(\lambda,\gamma/2)+\bar{\phi}'(\lambda,\gamma)
= -\rho(\lambda)-\frac{1}{N}\sum_{h=1}^{N_h}\delta(\lambda-\lambda_h)
\nonumber\\
+\int_{-\infty}^{\infty}d\lambda'\rho(\lambda')\phi'(\lambda-\lambda',\gamma) 
+ \frac{1}{N}\sum_{l=1}^{N_l}\bar{\phi}'(\lambda-z_l,\gamma).
\end{eqnarray}
Here we have introduced the new notations
\begin{equation}
\phi\left(\lambda,\frac{n\gamma}{2}\right)=\frac{1}{2\pi}f(\lambda,n,+1),
\qquad
\bar{\phi}\left(\lambda,\frac{n\gamma}{2}\right)=\frac{1}{2\pi}f(\lambda,n,-1)
\end{equation}
and the prime means the derivative with respect to the first argument.
The numbers of holes $\lambda_h$ and additional complex roots $z_l$ are denoted
by $N_h$ and $N_l$ respectively. The hole positions are defined as solutions of
(\ref{logbae}) for the omitted $I_h$.

The solution of (\ref{logbae1}) contains different contributions
obtained by Fourier transformation:
\begin{equation}
\rho(\lambda) = \rho_0(\lambda) + \frac{1}{N}(\rho_h(\lambda)+\rho_c(\lambda)
+\rho_w(\lambda))
\end{equation}
with
\begin{eqnarray}\label{fourier}
\hat{\rho_0}(p) &=& \frac{1+2\cosh(p\gamma/2)}{2\cosh(p(\pi-\gamma)/2)}
\nonumber\\
\hat{\rho_h}(p) &=& -\sum_{h=1}^{N_h} \frac{e^{-ip\lambda_h}\sinh
(p\pi/2)}{2\sinh(p\gamma/2)\cosh(p(\pi-\gamma)/2)}
\nonumber\\
\hat{\rho_c}(p) &=& -\sum_{l=1}^{N_c/2} e^{-ip\sigma_l} \frac{\cosh
(p\gamma/2)\cosh(p\tau_l)}{\cosh(p(\pi-\gamma)/2)}
\nonumber\\
\hat{\rho_w}(p) &=& \sum_{l=1}^{N_w/2} e^{-ip\sigma_l} \frac{\sinh(p
(\pi-\gamma)/2)\cosh(p(\pi-2\tau_l)/2)}{\cosh(p(\pi-\gamma)/2)\sinh(p\gamma/2)}
.
\end{eqnarray}
Here we have distinguished between complex close ($|Im(z_l)|<\pi/2-\gamma$) 
and wide roots ($|Im(z_l)|>\pi/2-\gamma$) differing in their Fourier 
transforms. Moreover we have used that complex roots appear in conjugated pairs
($z_l,\bar{z}_l$) only.

Energy and momentum of the elementary hole excitations are given through
\begin{eqnarray}
\varepsilon_h(\lambda_h) = \bar{\varepsilon}_h(\lambda_h) 
+ \tilde{\varepsilon}_h(\lambda_h),
\nonumber\\
\bar{\varepsilon}_h(\lambda_h) = - \frac{\pi\bar{c}}{\pi-\gamma}
\frac{1}{\cosh(\pi\lambda_h/(\pi-\gamma))},
\nonumber\\
\tilde{\varepsilon}_h(\lambda_h) = - \frac{4\pi\tilde{c}}{\pi-\gamma}
\frac{\cos(\pi\gamma/2(\pi-\gamma))\cosh(\pi\lambda_h/(\pi-\gamma))}
{\cosh(2\pi\lambda_h/(\pi-\gamma))+\cos(\pi\gamma/(\pi-\gamma))},
\end{eqnarray}
\begin{eqnarray}
\fl
p_h(\lambda_h) = \frac{1}{2}\arctan\left(\sinh\left(\frac{\pi\lambda_h}
{\pi-\gamma}\right)\right) + \frac{\pi}{4}
+ \arctan\left(\frac{\sinh(\pi\lambda_h/(\pi-\gamma))}
{\cos(\pi\gamma/2(\pi-\gamma))}\right)
+ \frac{\pi}{2}.
\end{eqnarray}

We did not find an explicit expression for the dispersion law of hole 
excitations. Nevertheless, numerical calculation suggests, that the curve shows
the expected behaviour (cf. figure 2).

Now we want to derive equations for hole positions and complex roots excluding 
the vacuum parameters by using the method  of Babelon {\it et al.}
\cite{babelon}. For this purpose we write down the BAE in integral 
approximation for a complex root $z$:
\begin{equation}\label{zbae}
\exp(2\pi i[I(z)-F(z)])=-1
\end{equation}
with
\begin{equation}\label{i}
I'(z) = N \int_{-\infty}^{\infty} d\lambda' \sigma(\lambda') \bar{\phi}'
(z-\lambda',\gamma),
\end{equation}
\begin{equation}
F'(z) = N \phi'(z,\frac{1}{2}\gamma) + N \phi'(z,\gamma) +
\sum_{h=1}^{N_h} \bar{\phi}'(z-\lambda_h,\gamma) +
\sum_{l=1}^{N_c} \phi'(z-z_l,\gamma),
\end{equation}
and $\sigma(\lambda)$ is the regular density
\begin{equation}
\sigma(\lambda)=\rho(\lambda)+\frac{1}{N}\sum_{h=1}^{N_h}\delta(\lambda-
\lambda_h).
\end{equation}
$I'(z)$ has discontinuities on the lines $Im(z)=\pm(\pi/2-\gamma$). So
analysing (\ref{zbae}) we have to consider three cases.

a)$\quad Im(z_l)>\pi/2-\gamma$.
We can evaluate the function $I(z)-F(z)$ directly by integrating
\begin{equation}\label{if}
I'(z)-F'(z) = - \sigma\left(z-\frac{i\pi}{2}\right)N.
\end{equation}
Noticing
\begin{equation}
I(\infty)-F(\infty) = 0
\end{equation}
one gets
\begin{equation}
I(z)-F(z) = - N \int_{\infty}^{z} \sigma\left(u-\frac{i\pi}{2}\right) du
\end{equation}
and
\begin{equation}\label{zbae1}
\exp\left(-2\pi i N \int_{\infty}^{z} \sigma\left(u-\frac{i\pi}{2}\right) du
\right) = -1.
\end{equation}
Splitting $\sigma(u)$ into vacuum and excitation contributions
\begin{equation}
\sigma\left(u-\frac{i\pi}{2}\right) = \sigma_0\left(u-\frac{i\pi}{2}\right)+
\frac{1}{N}\bar{\sigma}\left(u-\frac{i\pi}{2}\right)
\end{equation}
we get for (\ref{zbae1})
\begin{equation}
\exp\left(2\pi i N \int_{\infty}^{z} \sigma_0\left(u-\frac{i\pi}{2}\right)du +
2\pi i\int_{\infty}^{z}\bar{\sigma}\left(u-\frac{i\pi}{2}\right)du\right) = -1.
\end{equation}
The first exponent term contains an explicit $N$-dependence, while the 
second does not. So the validity of the above equation in the limit $N\to
\infty$ requires, that the second term cancels the $N$-dependence in the
real part of the first one by approaching $z$ in exponential order towards a 
pole of the integrand. These poles of $I'(u)-F'(u)$ are
\begin{equation}
z_l \pm i\gamma, \quad \lambda_h \pm (\frac{i\pi}{2}-i\gamma), \quad \pm
\frac{1}{2}i\gamma, \quad \pm i\gamma,
\end{equation}
with the relevant ones
\begin{equation}
z_l \pm i\gamma \qquad \mbox{with residue} \quad\mp i.
\end{equation}
Now it can be easily established that
\begin{equation}
Im \left(\int_{\infty}^{z} \sigma_0\left(u-\frac{i\pi}{2}\right)du\right)>0.
\end{equation}
Therefore it must exist $z_l$ with $z = z_l + i\gamma$.

b)$\quad -\pi/2+\gamma<Im(z_l)<\pi/2-\gamma$.
Here evaluating (\ref{i}) by deforming the integration contour provides an 
additional term due to residue theorem. One has to replace
\begin{equation}
I(z) \rightarrow I(z) - N \sigma\left(z-\frac{i\pi}{2}+i\gamma\right).
\end{equation}
Then (\ref{zbae}) reads
\begin{equation}\label{zbae2}
\exp\left(2\pi i N \int_{\infty}^{z} \left[\sigma\left(u-\frac{i\pi}{2}\right)
-\sigma\left(z-\frac{i\pi}{2}+i\gamma\right)\right] du\right) = -1.
\end{equation}
Using the Fourier transforms (\ref{fourier}) one can establish the relation
\begin{equation}\label{sig0per}
\sigma_0(u)+\sigma_0(u-i\pi+i\gamma) = 0.
\end{equation}
Then determining the sign of the real part of the exponent leads to
\begin{equation}
Im \left(\int_{\infty}^{z}\left[\sigma_0\left(u+\frac{i\pi}{2}\right)
+\sigma_0\left(u-\frac{i\pi}{2}\right)\right]du\right)
\left\{ \begin{array}{r@{\quad:\quad}l}>0 & Im(z)>0 \\ <0 & Im(z)<0\end{array}
\right.
\end{equation}
and we conclude that there exists $z_l$ with $z=z_l+i\gamma$ for $Im(z)>0$ and
$z_l$ with $z=z_l-i\gamma$ for $Im(z)<0$ respectively.

c)$\quad Im(z_l)<-\pi/2+\gamma$.
Evaluating (\ref{i}) again by contour deformation and substituting into 
(\ref{zbae}) leads to
\begin{equation}\label{zbae3}
\fl\exp\left(2\pi i N \int_{\infty}^{z} \left[\sigma\left(u-\frac{i\pi}{2}
\right)-\sigma\left(z-\frac{i\pi}{2}+i\gamma\right)+\sigma\left(z+\frac{i\pi}
{2}-i\gamma\right)\right] du\right) = -1.
\end{equation}
Using (\ref{sig0per}) the $N$-dependent exponent term reduces to
\begin{equation}
2\pi i N \int_{\infty}^{z} \sigma_0\left(u+\frac{i\pi}{2}\right)du.
\end{equation}
With
\begin{equation}
Im \left(\int_{\infty}^{z} \sigma_0\left(u+\frac{i\pi}{2}\right)du\right)<0
\end{equation}
follows $z = z_l - i\gamma$, where $z_l$ is a BAE root.

Now we are able to determine the configurations  in which complex roots can 
appear. We distinguish three cases for the anisotropy:

a)$\quad0<\gamma<\pi/4$.
Because of symmetry with respect to the real axis we only consider 
$Im(z)>0$. This leads to the existence of $z_l$ with $z=z_l+i\gamma$. 
Repeating this argument for $z_l$ ($Im(z_l)>0$ because of $\gamma<\pi/4$) 
generates another root $z_l-i\gamma$. This process generating roots breaks down
if the last generated root has negative imaginary part. Therefore the 'minimal'
configuration induced by $z$ is a chain with spacing $i\gamma$ situated 
completely in the upper half plane except the last generated root.

For further discussion one has to distinguish wide and close roots.
\begin{enumerate}
\item
$Im(z)>\pi/2-\gamma$. Taking the product over equations (\ref{zbae1}) and 
(\ref{zbae2}) respectively for all chain members shows that there remains a 
non-compensated $N$-dependent term until the chain is continued into the lower
region of wide roots. Then the configuration gets 'stable' in the limit 
$N\to\infty$. Notice that also the configuration with all roots complex 
conjugated exists, so that the configuration is actually a double chain.
\item
$0<Im(z)<\pi/2-\gamma$. Here the above procedure does not lead to a 'stable'
configuration. The only way for this configuration to exist in the limit 
$N\to\infty$ is arranging the roots symmetrically with respect to the real 
axis. Therefore this configuration is of the string type. It coincides with its
complex conjugated one (see figure 1).
\end{enumerate}

b)$\quad\pi/4<\gamma<\pi/3$. and c)$\quad\pi/3<\gamma<\pi/2$.
Analogous argumentations lead again to short double chains and strings 
(see figure 1).

\begin{figure}
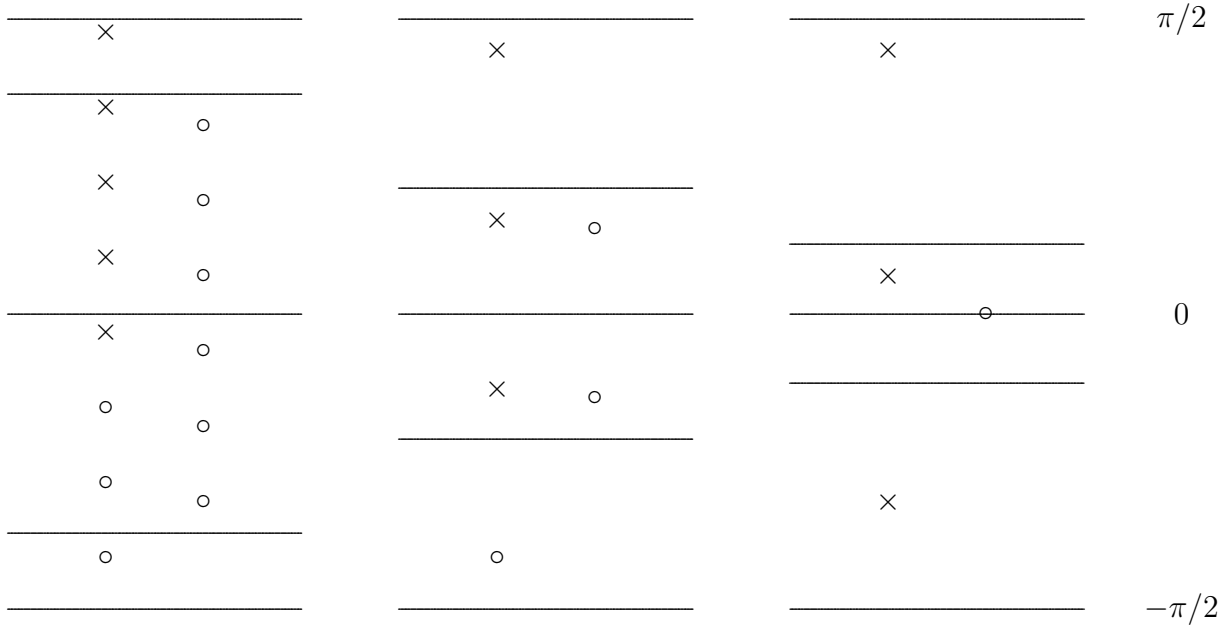

\input figure1.tex
\caption{Some special configurations of complex roots for the three cases of
$\gamma$. With '$\times$' a member of the 'minimal' configuration is symbolized
and '$\circ$' denotes additional roots to make the configuration 'stable'. 
String configuration members are denoted by '$\circ$' too due to the fact that
stability is realized by symmetrical arrangement. The lines varying in the 
three cases denotes $\pm(\pi/2-\gamma)$. Notice that for non-symmetric 
configurations the complex conjugated one in this picture is omitted.}
\end{figure}

We can write down the possible configurations for the three cases in a more 
compact way:
\begin{enumerate}
\item $z_t=z-it\gamma,t=0\dots n, n=\left[ \frac{\pi}{\gamma} \right], \left[ 
\frac{\pi}{\gamma} \right]-1$, $z$ is a wide root
\item $z_t=z-it\gamma,t=0\dots n,n\leq n_{max},n_{max}=\left[ \frac{\pi}
{\gamma}\right] - 2,Im(z)=\frac{n\gamma}{2}$, $z$ is a close root.
\end{enumerate}

The higher level BAE for a state with complex root configurations of the first
type can be obtained by taking the product over equations (\ref{zbae}) for all 
members and calculating $I(z)-F(z)$ directly via Fourier transforms. With the
new parameters
\begin{eqnarray}
\chi_l=z_l-\frac{in_l\gamma}{2},\quad\mbox{for $n$ even},\nonumber\\
\tilde{\chi}_l=z_l-\frac{in_l\gamma}{2},\quad\mbox{for $n$ odd},\nonumber\\
P\dots\mbox{number of $\chi$-configurations},\nonumber\\
\tilde{P}\dots\mbox{number of $\tilde{\chi}$-configurations},\nonumber\\
\alpha=\frac{\pi}{\pi-\gamma}
\end{eqnarray}
it follows
\begin{eqnarray}\label{hlbae}
\fl\prod_{h=1}^{N_h}\frac{\sinh(\alpha(\chi_l-\lambda_h)+i\alpha\pi/2)}{\sinh
(\alpha(\chi_l-\lambda_h)-i\alpha\pi/2)}=\!-\!\!\prod_{j=1}^{P}\!
\frac{\sinh(\alpha(\chi_l-\chi_j)+i\alpha\pi)}{\sinh(\alpha(\chi_l-\chi_j)
-i\alpha\pi)}
\!\prod_{j=1}^{\tilde{P}}\!\frac{\cosh(\alpha(\chi_l-\tilde{\chi}_j)+i\alpha
\pi)}{\cosh(\alpha(\chi_l-\tilde{\chi}_j)-i\alpha\pi)},
\nonumber\\
\fl\prod_{h=1}^{N_h}\frac{\cosh(\alpha(\tilde{\chi}_l-\lambda_h)+i\alpha\pi/2)}
{\cosh(\alpha(\tilde{\chi}_l-\lambda_h)-i\alpha\pi/2)}=\!-\!\!\prod_{j=1}^{P}\!
\frac{\cosh(\alpha(\tilde{\chi}_l-\chi_j)+i\alpha\pi)}{\cosh(\alpha(\tilde{\chi
}_l-\chi_j)-i\alpha\pi)}\!\prod_{j=1}^{\tilde{P}}\!\frac{\sinh(\alpha(\tilde
{\chi}_l-\tilde{\chi}_j)+i\alpha\pi)}{\sinh(\alpha(\tilde{\chi}_l-\tilde
{\chi}_j)-i\alpha\pi)}.
\nonumber\\
\end{eqnarray}
Calculating energy and momentum for the configurations shows, that for the 
first type the direct contribution to these magnitudes vanishes. On the other
hand there are two hole parameters associated to such a configuration for 
symmetrical arrangement ('string type') and four for a non-symmetric 
configuration ('double chain'). So these configurations do carry energy, but 
only via their associated holes. This is the standard picture, which holds also
in our model.

For the second type we get
\begin{eqnarray}\label{ex2en}
\fl \Delta E = -\bar{c}\frac{4\pi}{\pi-\gamma}\frac{\sin\left[(n+1)\gamma
\pi/2(\pi-\gamma)\right]\cosh\left[ \sigma\pi/(\pi-\gamma)\right]}{\cosh\left[ 
2\sigma\pi/(\pi-\gamma)\right]-\cos\left[ (n+1)\gamma\pi/(\pi-\gamma)\right]}
\nonumber\\
-\tilde{c}\frac{4\pi}{\pi-\gamma}\left\{ \frac{\sin\left[(n+2)\gamma\pi/2(\pi-
\gamma)\right]\cosh\left[ \sigma\pi/(\pi-\gamma)\right]}{\cosh\left[ 2\sigma\pi
/(\pi-\gamma)\right]-\cos\left[ (n+2)\gamma\pi/(\pi-\gamma) \right]} 
\right.
\nonumber\\
\left.
+ \frac{\sin\left[n\gamma\pi/2(\pi-\gamma)\right]\cosh\left[ \sigma\pi/(\pi-
\gamma)\right]}{\cosh\left[ 2\sigma\pi/(\pi-\gamma)\right]-\cos\left[ n\gamma
\pi/(\pi-\gamma)\right]} \right\}
\end{eqnarray}
\begin{eqnarray}\label{ex2mom}
\fl\Delta P = \sum_{i=0}^2 \arctan \left( \frac{\sinh \left[ \sigma\pi/(\pi-
\gamma)\right]}{\sin\left[(n+i)\gamma\pi/2(\pi-\gamma) \right]}\right)
+\frac{3\pi}{2}
\end{eqnarray}
for $n\geq 1$. The dispersion law of these excitations has 
been calculated numerically for different $\gamma<\pi/3$ and string length $n$ 
and $\bar{c}/\tilde{c}=1$. A comparison with the dispersion law of a free two 
magnon state, where the momenta of the magnons are equal as given in figure 2 
reveals, that these excitations can be identified with bound magnon states in 
analogy to \cite{babelon}. One can see from the picture, that the maximum of 
the curve decreases monotonically with increasing $n$ up to $n_{max}$ for 
$\gamma$ fix, approaching the dispersion curve for the free state. On the other
hand for $\gamma\to\pi/3$, where only the $n=1$ state exists, again the bound 
state curve approaches the one for the free state monotonically.

Moreover there is a state with $n=0$ allowed through the whole $\gamma$-range.
Its energy and momentum are given via (\ref{ex2en}) and (\ref{ex2mom}), where 
in the last formula the term for $i=0$ is dropped. This leads to a range for 
the momentum between those for a single hole excitation and a bound state. 
Therefore the bound state interpretation fails.

\begin{figure}
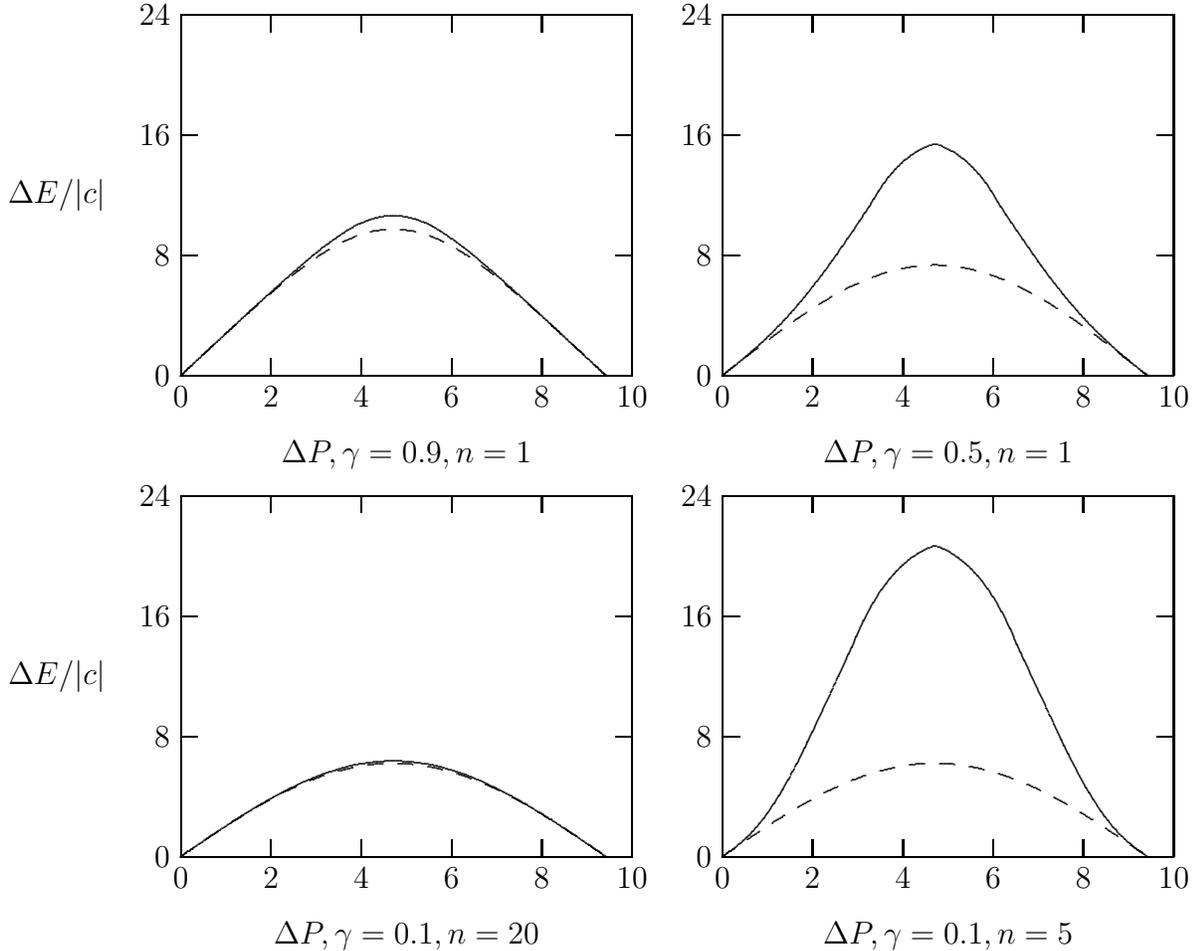

\input figure2.tex
\caption{Dispersion relation for Type II excitations (solid line) compared with
the dispersion law for a free two magnon state (dotted line) for different 
$\gamma$.}
\end{figure}

\section{Conclusions}
We have investigated a generalized Heisenberg spin chain with alternating spins
XXZ($\frac{1}{2},1$) in the gapless region. By means of thermodynamic Bethe 
ansatz (TBA) integral equations determining the ground state have been derived.
In the case of negative coupling of spin interactions these equations were 
solved. It turns out that the ground state is formed by a sea of $(1,-)$-
strings and is therefore of antiferromagnetic nature.

Weakly excited states above this antiferromagnetic vacuum have been analysed
following the method introduced in \cite{babelon}. In analogy to the XXZ
($\frac{1}{2}$) model for $-1<\Delta<0$ two types of excitations appear. The 
first one are scattering states of magnons. Higher level BAE are derived which 
determine the parameters of these excitations. The second type can be 
identified with bound magnon states in analogy to \cite{babelon}.

\section*{Acknowledgment}
We would like to thank H. M. Babujian for helpful discussions.

\newpage
\section*{Figure captions}
{\bf Figure 1.} Some special configurations of complex roots for the three 
cases of $\gamma$. With '$\times$' a member of the 'minimal' configuration is 
symbolized and '$\circ$' denotes additional roots to make the configuration 
'stable'. String configuration members are denoted by '$\circ$' too due to the 
fact that stability is realized by symmetrical arrangement. The lines varying 
in the three cases denotes $\pm(\pi/2-\gamma)$. Notice that for non-symmetric 
configurations the complex conjugated one in this picture is omitted.\\[2cm]
{\bf Figure 2.} Dispersion relation for Type II excitations (solid line) 
compared with the dispersion law for a free two magnon state (dotted line) for 
different $\gamma$.

\end{document}